\documentclass[pra,showpacs,showkeys,amsfonts]{revtex4}
\usepackage{graphicx}
\begin{document}

\def\ttt{{\rm Tr} }
\def\diag{{\rm diag} }

\title{Quantum information in base $n$ defined by state partitions}
\author{Karl Svozil}
 \email{svozil@tuwien.ac.at}
\homepage{http://tph.tuwien.ac.at/~svozil}
\affiliation{Institut f\"ur Theoretische Physik, University of Technology Vienna,
Wiedner Hauptstra\ss e 8-10/136, A-1040 Vienna, Austria}

\begin{abstract}
We define a ``nit'' as a radix $n$ measure of quantum information which
is based on state partitions associated with the outcomes of
$n$-ary observables and which, for $n>2$, is fundamentally irreducible to a binary coding.
Properties of this measure for  entangled many-particle states are discussed.
$k$ particles specify $k$ nits in such a way that $k$ mutually commuting
measurements of observables with $n$ possible outcomes
are sufficient to determine the information.
\end{abstract}


\maketitle

The formal concept of  information is tied to physics,
at least as far as applicability is a concern.
There seems to be one issue,
which, despite notable exceptions (e.g., \cite[Footnote 6]{zeil-99} and
\cite{Muthukrishnan}),
has not yet been acknowledged widely:
the principal irreducibility of quantum information in base $n$.
Define a ``nit'' as a unit of information equal to the
amount of information obtained by learning which of $n$ equally likely events occurred.
An $n$-state particle  can be prepared in a single one of $n$ possible states.
Then, this particle carries one nit of information, namely to
``be in a single one from $n$ different states.''
Subsequent measurements may confirm this statement.
The most natural code basis for such a configuration is $n$,
and not a binary  one.

Classically, there is no preferred code basis whatsoever.
Every classical state is postulated to be determined by a
point in phase space.
Formally, this amounts to an infinite amount of information in whatever base,
since
with probability one, all points are random; i.e., algorithmically incompressible
\cite{calude:94}.
Operationally, only a finite amount of classical information is accessible.
Yet, the particular base in which this
finite amount of classical information is represented
is purely conventional.
The same holds true for discrete classical systems,
such as $n$ modes of vibration on a string, where the restriction
to these particular states is rather arbitrary.

The fundamental difference between classical and quantum information
with respect to code bases
could be illustrated by the following example.
Physically, each nit could be
represented by an $n$-level system.
A single measurement collapses an $n$-state
superposition and yields only one output, not $\log_2 n$ outputs.
In the nonentangled $k$ particle case,
the $k$ mutually commuting observables
could be some physical quantity (e.g., energy levels) associated with each
particle.
This sets the stage for the more general observables
associated with ``entangled'' states.
References \cite{zeil-99} and \cite{DonSvo01} discuss
examples with Bell states and Greenberger-Horne-Zeilinger states
for the binary case, respectively.

In what follows, let us always consider a complete
system of base states  ${\cal B}$ associated with a unique ``context''
or ``communication frame''
 ${\cal F}=\{F_1,F_2, \ldots ,F_{k} \}$, which corresponds to
co-measurable observables with $n$ outcomes.
For $n=2$,
their explicit form has been enumerated in  \cite{DonSvo01}.
In this particular case, the $F$'s can be identified with certain projection
operators from the set of all possible mutually orthogonal ones,
whose two eigenvalues can be identified with the two states.
For three or more particles, this is no longer possible.

It should be emphasized that only the case of an entanglement between different particles
but not within each particle is considered.
If more than one observable could be associated with each particle,
then these can become entangled as well, and then $k$ $n$-ary observables will no longer
be sufficient to describe $k$ particles.


For a single $n$-state particle, the nit can be formalized as a state partition
which is fine grained into $n$ elements, one state per element.
That is,
if the set of states is represented by $\{1,\ldots ,n\}$,
then the nit is defined by
$
\{\{1\} ,\ldots ,\{n\}\}.
$
Of course, any labeling would suffice, as long as the structure is preserved.
It does not matter
whether one calls the states, for instance, ``+,'' ``0'' and  ``-'', or ``1,'' ``2'' and ``3'',
resulting in a trit represented by
$\{\{+\} ,\{0\} ,\{-\}\}$ or
$\{\{1\} ,\{2\} ,\{3\}\}$
(here, the term ``trit'' stands for a nit with $n=3$).
Thus, nits are defined modulo isomorphisms (i.e., one-to-one translations)
of the state labels.
To complete the setup of the single particle case, let us
recall that any such state set would correspond to an orthonormal
basis of $n$-dimensional Hilbert space.

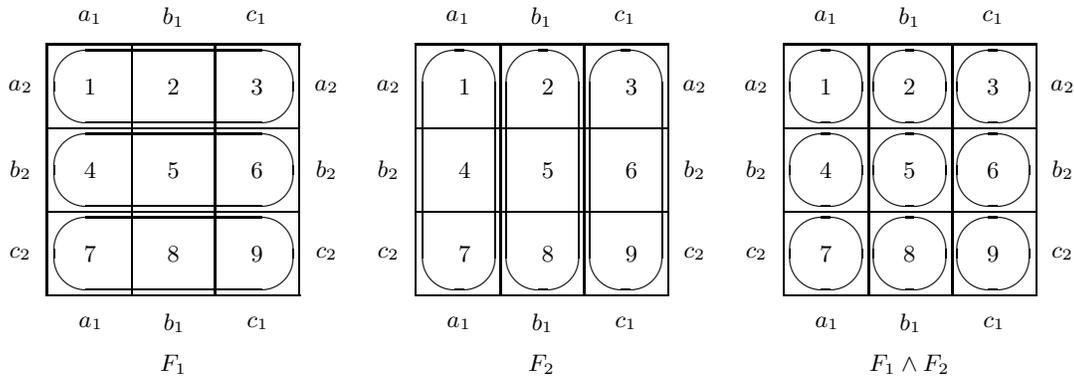
\begin{figure}
\begin{tabular}{ccccccc}
\unitlength 0.37mm
\linethickness{0.4pt}
\begin{picture}(110.00,125.00)
\put(9.67,25.00){\framebox(90.33,90.00)[cc]{}}
\put(40.00,115.00){\line(0,-1){90.00}}
\put(70.00,115.00){\line(0,-1){90.00}}
\put(9.67,55.00){\line(1,0){90.33}}
\put(100.00,85.00){\line(-1,0){90.33}}
\put(0.00,40.00){\makebox(0,0)[cc]{$c_2$}}
\put(0.00,70.00){\makebox(0,0)[cc]{$b_2$}}
\put(0.00,100.00){\makebox(0,0)[cc]{$a_2$}}
\put(25.00,125.00){\makebox(0,0)[cc]{$a_1$}}
\put(55.00,125.00){\makebox(0,0)[cc]{$b_1$}}
\put(85.00,125.00){\makebox(0,0)[cc]{$c_1$}}
\put(25.33,15.00){\makebox(0,0)[cc]{$a_1$}}
\put(55.33,15.00){\makebox(0,0)[cc]{$b_1$}}
\put(85.33,15.00){\makebox(0,0)[cc]{$c_1$}}
\put(110.00,40.00){\makebox(0,0)[cc]{$c_2$}}
\put(110.00,70.00){\makebox(0,0)[cc]{$b_2$}}
\put(110.00,100.00){\makebox(0,0)[cc]{$a_2$}}
\put(55.00,100.00){\oval(86.00,26.00)[]}
\put(25.00,100.00){\makebox(0,0)[cc]{$1$}}
\put(55.00,100.00){\makebox(0,0)[cc]{$2$}}
\put(85.00,100.00){\makebox(0,0)[cc]{$3$}}
\put(55.00,70.00){\oval(86.00,26.00)[]}
\put(55.00,40.00){\oval(86.00,26.00)[]}
\put(25.00,70.00){\makebox(0,0)[cc]{$4$}}
\put(55.00,70.00){\makebox(0,0)[cc]{$5$}}
\put(85.00,70.00){\makebox(0,0)[cc]{$6$}}
\put(25.00,40.00){\makebox(0,0)[cc]{$7$}}
\put(55.00,40.00){\makebox(0,0)[cc]{$8$}}
\put(85.00,40.00){\makebox(0,0)[cc]{$9$}}
\put(55.00,0.00){\makebox(0,0)[cc]{$F_1$}}
\end{picture}
&
\quad
\unitlength 0.50mm
\linethickness{0.4pt}
\begin{picture}(0.00,105.00)
\put(0.00,52.00){\makebox(0,0)[cc]{$\;$}}
\end{picture}
&
\unitlength 0.37mm
\linethickness{0.4pt}
\begin{picture}(110.00,120.00)
\put(10.00,25.00){\framebox(90.33,90.00)[cc]{}}
\put(40.33,115.00){\line(0,-1){90.00}}
\put(70.33,115.00){\line(0,-1){90.00}}
\put(10.00,55.00){\line(1,0){90.33}}
\put(100.33,85.00){\line(-1,0){90.33}}
\put(0.00,40.00){\makebox(0,0)[cc]{$c_2$}}
\put(0.00,70.00){\makebox(0,0)[cc]{$b_2$}}
\put(0.00,100.00){\makebox(0,0)[cc]{$a_2$}}
\put(25.00,125.00){\makebox(0,0)[cc]{$a_1$}}
\put(55.00,125.00){\makebox(0,0)[cc]{$b_1$}}
\put(85.00,125.00){\makebox(0,0)[cc]{$c_1$}}
\put(25.33,15.00){\makebox(0,0)[cc]{$a_1$}}
\put(55.33,15.00){\makebox(0,0)[cc]{$b_1$}}
\put(85.33,15.00){\makebox(0,0)[cc]{$c_1$}}
\put(110.00,40.00){\makebox(0,0)[cc]{$c_2$}}
\put(110.00,70.00){\makebox(0,0)[cc]{$b_2$}}
\put(110.00,100.00){\makebox(0,0)[cc]{$a_2$}}
\put(25.33,70.00){\oval(26.00,86.00)[]}
\put(55.33,70.00){\oval(26.00,86.00)[]}
\put(85.33,70.00){\oval(26.00,86.00)[]}
\put(27.33,100.00){\makebox(0,0)[cc]{$1$}}
\put(27.33,70.00){\makebox(0,0)[cc]{$4$}}
\put(27.33,40.00){\makebox(0,0)[cc]{$7$}}
\put(57.33,100.00){\makebox(0,0)[cc]{$2$}}
\put(57.33,70.00){\makebox(0,0)[cc]{$5$}}
\put(57.33,40.00){\makebox(0,0)[cc]{$8$}}
\put(87.33,100.00){\makebox(0,0)[cc]{$3$}}
\put(87.33,70.00){\makebox(0,0)[cc]{$6$}}
\put(87.33,40.00){\makebox(0,0)[cc]{$9$}}
\put(55.00,0.00){\makebox(0,0)[cc]{$F_2$}}
\end{picture}
&
\quad
\unitlength 0.50mm
\linethickness{0.4pt}
\begin{picture}(0.00,105.00)
\put(0.00,52.00){\makebox(0,0)[cc]{$\;$}}
\end{picture}
&
\unitlength 0.37mm
\linethickness{0.4pt}
\begin{picture}(110.0,120.00)
\put(10.00,25.00){\framebox(90.33,90.00)[cc]{}}
\put(40.33,115.00){\line(0,-1){90.00}}
\put(70.33,115.00){\line(0,-1){90.00}}
\put(10.00,55.00){\line(1,0){90.33}}
\put(100.33,85.00){\line(-1,0){90.33}}
\put(0.00,40.00){\makebox(0,0)[cc]{$c_2$}}
\put(0.00,70.00){\makebox(0,0)[cc]{$b_2$}}
\put(0.00,100.00){\makebox(0,0)[cc]{$a_2$}}
\put(25.00,125.00){\makebox(0,0)[cc]{$a_1$}}
\put(55.00,125.00){\makebox(0,0)[cc]{$b_1$}}
\put(85.00,125.00){\makebox(0,0)[cc]{$c_1$}}
\put(25.33,15.00){\makebox(0,0)[cc]{$a_1$}}
\put(55.33,15.00){\makebox(0,0)[cc]{$b_1$}}
\put(85.33,15.00){\makebox(0,0)[cc]{$c_1$}}
\put(110.00,40.00){\makebox(0,0)[cc]{$c_2$}}
\put(110.00,70.00){\makebox(0,0)[cc]{$b_2$}}
\put(110.00,100.00){\makebox(0,0)[cc]{$a_2$}}
\put(25.00,100.00){\oval(26.00,26.00)[]}
\put(25.00,100.00){\makebox(0,0)[cc]{1}}
\put(55.00,100.00){\oval(26.00,26.00)[]}
\put(85.00,100.00){\oval(26.00,26.00)[]}
\put(55.00,100.00){\makebox(0,0)[cc]{2}}
\put(85.00,100.00){\makebox(0,0)[cc]{3}}
\put(25.00,70.00){\oval(26.00,26.00)[]}
\put(25.00,40.00){\oval(26.00,26.00)[]}
\put(25.00,70.00){\makebox(0,0)[cc]{4}}
\put(25.00,40.00){\makebox(0,0)[cc]{7}}
\put(55.00,70.00){\oval(26.00,26.00)[]}
\put(55.00,40.00){\oval(26.00,26.00)[]}
\put(85.00,70.00){\oval(26.00,26.00)[]}
\put(85.00,40.00){\oval(26.00,26.00)[]}
\put(55.00,70.00){\makebox(0,0)[cc]{5}}
\put(55.00,40.00){\makebox(0,0)[cc]{8}}
\put(85.00,70.00){\makebox(0,0)[cc]{6}}
\put(85.00,40.00){\makebox(0,0)[cc]{9}}
\put(55.00,0.00){\makebox(0,0)[cc]{$F_1\wedge F_2$}}
\end{picture}
\end{tabular}
 \caption{
Representation of
state partitions as cells of a two-dimensional square spanned
by the single cells of the two three-state particles.
}
\label{2002-statepart1}
\end{figure}

Before proceeding to the most general case,
we shall consider the case of two particles
with three states per particle in all details.
We shall adopt an $n$-ary search strategy.
Assume that the first and the second particle
has three orthogonal states labeled by
$a_1,b_1,c_1$
and
$a_2,b_2,c_2$,
respectively.
Then nine product states can be formed and labeled from $1$ to $9$ in
lexicographic order; i.e.,
$
a_1a_2 \equiv 1,   \cdots , c_1c_2 \equiv 9
$.
Consider a set of two
comeasurable three-valued observables inducing two state partitions
of the set of states $S=\{1,2,\ldots , 9\}$ with three partition
elements with the properties
that (i) the set theoretic intersection of any two elements of the two
partitions is a single state, and (ii) the union of all these nine
intersections is just the set of state $S$.
As can be easily checked, an example for such state partitions are
\begin{equation}
\begin{array}{llllll}
F_1&=&\{\{1,2,3\},\{4,5,6\},\{7,8,9\}\}&\equiv& \{\{a_1\},\{b_1\},\{c_1\}\},\\
F_2&=&\{\{1,4,7\},\{2,5,8\},\{3,6,9\}\}&\equiv& \{\{a_2\},\{b_2\},\{c_2\}\}.\\
\end{array}
\label{2002-statepart-ps3e}
\end{equation}
Operationally, the trit $F_1$ can be obtained by
measuring the first particle state:
$\{1,2,3\}$ is associated with state $a_1$,
$\{4,5,6\}$ is associated with $b_1$,  and
$\{7,8,9\}$ is associated with $c_1$.
The trit $F_2$ can be obtained by
measuring the state of the second particle:
$\{1,4,7\}$ is associated with state $a_2$,
$\{2,5,8\}$ is associated with $b_2$,  and
$\{3,6,9\}$ is associated with $c_2$.
This amounts to the operationalization of the trits (\ref{2002-statepart-ps3e})
as state filters.
In the above case, the filters are ``local'' and can be realized on single particles,
one trit per particle.
In the more general case of rotated ``entangled'' states (cf. below),
the trits (more generally, nits) become
inevitably associated with joint properties of ensembles of particles.
Measurement of the propositions,
{\em ``the particle is in state $\{1,2,3\}$''}
and,
{\em ``the particle is in state $\{3,6,9\}$''}
can be evaluated by taking the set theoretic intersection of the respective sets; i.e., by
the proposition,
{\em ``the particle is in state $\{1,2,3\}\cap \{3,6,9\} = 3$.''}
In figure \ref{2002-statepart1},
the state partitions are drawn as cells of a two-dimensional square spanned
by the single cells of the two three-state particles.

A Hilbert space representation of this setting can be obtained as
follows.
Define the states in $S$ to be one-dimensional linear subspaces of
nine-dimensional Hilbert space; e.g.,
$
1 \equiv (1,0,0,0,0,0,0,0,0),\cdots ,
9 \equiv (0,0,0,0,0,0,0,0,1).
$
The trit operators are given by (the terms ``trit operator,'' ``observable,'' and the
corresponding state partition will be used synonymously)
\begin{equation}
\begin{array}{llll}
F_1&=& \diag (a,a,a,b,b,b,c,c,c),\\
F_2&=& \diag (a,b,c,a,b,c,a,b,c),\\
\end{array}
\label{2002-statepart-ps3top}
\end{equation}
for $a,b,c \in {\Bbb R}$, $a\neq b\neq c\neq a$.

If $F_2= \diag (d,e,f,d,e,f,d,e,f)$
and $a,b,c,d,e,f,$ are six different prime numbers,
then, due to the uniqueness of prime decompositions,
the two trit operators
can be combined to a single
``context'' operator
\begin{equation}
C=F_1\cdot F_2=F_2\cdot F_1=
\diag (ad,ae,af,bd,be,bf,cd,ce,cf)
\label{2002-statepart-ps3pd}
\end{equation}
which acts on both particles and has nine different eigenvalues.
Just as for the two-particle case \cite{DonSvo01},
there exist $3^2!=9!=362880$ permutations of operators
which are all able to separate the nine states.
They are obtained by forming a $(2\times 9)$-matrix
whose rows are the diagonal components of $F_1$ and $F_2$
from Eq. (\ref{2002-statepart-ps3top})
and permuting all the columns.
The resulting new operators $F_1'$ and $F_2'$ are also trit operators.

A generalization to $k$ particles in $n$ states per particle is straightforward.
We obtain
$k$ partitions of the product states with
$n$ elements per partition in such a way that
every single product state is obtained by the set theoretic intersection of
$k$ elements of all the different partitions.

Every single such partition can be interpreted as a nit.
All such sets are generated by permuting the set of states,
which amounts to $n^k!$ equivalent sets of state partitions.
However, since they are mere one-to-one translations,
they represent the same nits.
This equivalence, however, does not concern the property of (non)entanglement,
since the permutations take entangled states into nonentangled ones.
We shall give an example below.

Again, the standard orthonormal basis of
$n^k$-dimensional Hilbert space is identified with the set of states $S=\{1,2,\ldots ,n^k\}$; i.e.,
(superscript ``$T$'' indicates transposition)
\begin{equation}
\begin{array}{llll}
1 &\equiv& (1,\ldots,0)^T\equiv \mid 1,\ldots ,1\rangle = \mid 1\rangle \otimes \cdots \otimes \mid 1\rangle ,\\
  &\vdots&\\
n^k &\equiv& (0,\ldots,1)^T\equiv \mid n,\ldots ,n\rangle = \mid n\rangle \otimes \cdots \otimes \mid n\rangle .\\
\end{array}
\label{2002-statepart-psma}
\end{equation}
The single-particle states are also labeled by $1$ through $n$,
and the tensor product states are formed and ordered lexicographically ($0<1$).

The nit operators are defined via diagonal matrices
which contain equal amounts $n^{k-1}$ of mutually $n$ different numbers
such as different primes $q_1,\ldots ,q_n$; i.e.,
\begin{equation}
\begin{array}{llll}
F_1&=& \diag (\underbrace{\underbrace{q_1,\ldots ,q_1}_{n^{k-1}\;{\rm times}},\ldots ,\underbrace{q_n,\ldots ,q_n}_{n^{k-1}\;{\rm times}}}_{n^0\;{\rm times}}),\\
F_2&=& \diag (\underbrace{\underbrace{q_1,\ldots ,q_1}_{n^{k-2}\;{\rm times}},\ldots ,\underbrace{q_n,\ldots ,q_n}_{n^{k-2}\;{\rm times}}}_{n^1\;{\rm times}}),\\
  &\vdots&\\
F_k&=& \diag (\underbrace{q_1,\ldots ,q_n}_{n^{k-1}\;{\rm times}}).
\end{array}
\label{2002-statepart-nitopgen}
\end{equation}
The operators implement an $n$-ary search strategy,
filtering the search space into $n$ equal partitions of states,
such that a successive applications of all such filters
renders a single state.

There exist $n^k!$  sets of nit operators,
which are
are obtained by forming a $(k \times n^k)$-matrix
whose rows are the diagonal components of $F_1,\ldots,F_k$  from Eq.
(\ref{2002-statepart-nitopgen})
and permuting all the columns.
The resulting new operators $F_1',\ldots,F_k'$ are also nit operators.

All partitions discussed so far are equally weighted and well balanced,
as all elements of them contain an equal number of states.
In principle, one could also consider nonbalanced partitions.
For example, one could take the partition
$\overline{F}_1=\{\{1\},\{2,3\},\{4,5,6,7,8,9\}\}$
instead of $F_1$ in (\ref{2002-statepart-ps3e}),
represented the by trit diagonal operator
$\diag (a,b,b,c,c,c,c,c,c)$.
Yet any such attempt would result
in a deviation from the optimal $n$-ary search strategy, and
in an nonoptimal measurement procedures.
Another, more principal, disadvantage would be the fact that such a state separation
could not reflect the inevitable $n$-arity of the quantum choice.


%

\begin{figure}
\begin{tabular}{ccccccc}
\unitlength 0.37mm
\linethickness{0.4pt}
\begin{picture}(110.00,125.00)
\put(9.67,25.00){\framebox(90.33,90.00)[cc]{}}
\put(40.00,115.00){\line(0,-1){90.00}}
\put(70.00,115.00){\line(0,-1){90.00}}
\put(9.67,55.00){\line(1,0){90.33}}
\put(100.00,85.00){\line(-1,0){90.33}}
\put(0.00,40.00){\makebox(0,0)[cc]{$c_2$}}
\put(0.00,70.00){\makebox(0,0)[cc]{$b_2$}}
\put(0.00,100.00){\makebox(0,0)[cc]{$a_2$}}
\put(25.00,125.00){\makebox(0,0)[cc]{$a_1$}}
\put(55.00,125.00){\makebox(0,0)[cc]{$b_1$}}
\put(85.00,125.00){\makebox(0,0)[cc]{$c_1$}}
\put(25.33,15.00){\makebox(0,0)[cc]{$a_1$}}
\put(55.33,15.00){\makebox(0,0)[cc]{$b_1$}}
\put(85.33,15.00){\makebox(0,0)[cc]{$c_1$}}
\put(110.00,40.00){\makebox(0,0)[cc]{$c_2$}}
\put(110.00,70.00){\makebox(0,0)[cc]{$b_2$}}
\put(110.00,100.00){\makebox(0,0)[cc]{$a_2$}}
\put(12.00,113.00){\line(0,-1){13.00}}
\put(12.00,100.00){\line(1,-1){73.00}}
\put(85.00,27.00){\line(1,0){13.00}}
\put(98.00,27.00){\line(0,1){13.00}}
\put(98.00,40.00){\line(-1,1){73.00}}
\put(25.00,113.00){\line(-1,0){13.00}}
\put(12.00,83.00){\line(0,-1){13.00}}
\put(12.00,53.00){\line(0,-1){13.00}}
\put(25.00,83.00){\line(-1,0){13.00}}
\put(25.00,53.00){\line(-1,0){13.00}}
\put(55.00,27.00){\line(1,0){13.00}}
\put(25.00,27.00){\line(1,0){13.00}}
\put(68.00,27.00){\line(0,1){13.00}}
\put(38.00,27.00){\line(0,1){13.00}}
\put(38.00,40.00){\line(-1,1){13.00}}
\put(55.00,27.00){\line(-1,1){43.00}}
\put(68.00,40.00){\line(-1,1){43.00}}
\put(25.00,27.00){\line(-1,1){13.00}}
\put(42.00,113.00){\line(1,0){13.00}}
\put(72.00,113.00){\line(1,0){13.00}}
\put(42.00,100.00){\line(0,1){13.00}}
\put(72.00,100.00){\line(0,1){13.00}}
\put(98.00,70.00){\line(0,-1){13.00}}
\put(98.00,100.00){\line(0,-1){13.00}}
\put(98.00,57.00){\line(-1,0){13.00}}
\put(98.00,87.00){\line(-1,0){13.00}}
\put(85.00,87.00){\line(-1,1){13.00}}
\put(98.00,70.00){\line(-1,1){43.00}}
\put(85.00,57.00){\line(-1,1){43.00}}
\put(98.00,100.00){\line(-1,1){13.00}}
\put(25.00,100.00){\makebox(0,0)[cc]{1}}
\put(55.00,100.00){\makebox(0,0)[cc]{2}}
\put(85.00,100.00){\makebox(0,0)[cc]{3}}
\put(25.00,70.00){\makebox(0,0)[cc]{4}}
\put(25.00,40.00){\makebox(0,0)[cc]{7}}
\put(55.00,70.00){\makebox(0,0)[cc]{5}}
\put(55.00,40.00){\makebox(0,0)[cc]{8}}
\put(85.00,70.00){\makebox(0,0)[cc]{6}}
\put(85.00,40.00){\makebox(0,0)[cc]{9}}
\put(55.00,0.00){\makebox(0,0)[cc]{$F_1$}}
\end{picture}
&
\quad
\unitlength 0.50mm
\linethickness{0.4pt}
\begin{picture}(0.00,105.00)
\put(0.00,52.00){\makebox(0,0)[cc]{$\;$}}
\end{picture}
&
\unitlength 0.37mm
\linethickness{0.4pt}
\begin{picture}(110.00,125.00)
\put(9.67,25.00){\framebox(90.33,90.00)[cc]{}}
\put(40.00,115.00){\line(0,-1){90.00}}
\put(70.00,115.00){\line(0,-1){90.00}}
\put(9.67,55.00){\line(1,0){90.33}}
\put(100.00,85.00){\line(-1,0){90.33}}
\put(0.00,40.00){\makebox(0,0)[cc]{$c_2$}}
\put(0.00,70.00){\makebox(0,0)[cc]{$b_2$}}
\put(0.00,100.00){\makebox(0,0)[cc]{$a_2$}}
\put(25.00,125.00){\makebox(0,0)[cc]{$a_1$}}
\put(55.00,125.00){\makebox(0,0)[cc]{$b_1$}}
\put(85.00,125.00){\makebox(0,0)[cc]{$c_1$}}
\put(25.33,15.00){\makebox(0,0)[cc]{$a_1$}}
\put(55.33,15.00){\makebox(0,0)[cc]{$b_1$}}
\put(85.33,15.00){\makebox(0,0)[cc]{$c_1$}}
\put(110.00,40.00){\makebox(0,0)[cc]{$c_2$}}
\put(110.00,70.00){\makebox(0,0)[cc]{$b_2$}}
\put(110.00,100.00){\makebox(0,0)[cc]{$a_2$}}
\put(98.00,113.00){\line(-1,0){13.00}}
\put(85.00,113.00){\line(-1,-1){73.00}}
\put(12.00,40.00){\line(0,-1){13.00}}
\put(12.00,27.00){\line(1,0){13.00}}
\put(25.00,27.00){\line(1,1){73.00}}
\put(98.00,100.00){\line(0,1){13.00}}
\put(68.00,113.00){\line(-1,0){13.00}}
\put(38.00,113.00){\line(-1,0){13.00}}
\put(68.00,100.00){\line(0,1){13.00}}
\put(38.00,100.00){\line(0,1){13.00}}
\put(12.00,70.00){\line(0,-1){13.00}}
\put(12.00,100.00){\line(0,-1){13.00}}
\put(12.00,57.00){\line(1,0){13.00}}
\put(12.00,87.00){\line(1,0){13.00}}
\put(25.00,87.00){\line(1,1){13.00}}
\put(12.00,70.00){\line(1,1){43.00}}
\put(25.00,57.00){\line(1,1){43.00}}
\put(12.00,100.00){\line(1,1){13.00}}
\put(98.00,83.00){\line(0,-1){13.00}}
\put(98.00,53.00){\line(0,-1){13.00}}
\put(85.00,83.00){\line(1,0){13.00}}
\put(85.00,53.00){\line(1,0){13.00}}
\put(55.00,27.00){\line(-1,0){13.00}}
\put(85.00,27.00){\line(-1,0){13.00}}
\put(42.00,27.00){\line(0,1){13.00}}
\put(72.00,27.00){\line(0,1){13.00}}
\put(72.00,40.00){\line(1,1){13.00}}
\put(55.00,27.00){\line(1,1){43.00}}
\put(42.00,40.00){\line(1,1){43.00}}
\put(85.00,27.00){\line(1,1){13.00}}
\put(25.00,100.00){\makebox(0,0)[cc]{1}}
\put(55.00,100.00){\makebox(0,0)[cc]{2}}
\put(85.00,100.00){\makebox(0,0)[cc]{3}}
\put(25.00,70.00){\makebox(0,0)[cc]{4}}
\put(25.00,40.00){\makebox(0,0)[cc]{7}}
\put(55.00,70.00){\makebox(0,0)[cc]{5}}
\put(55.00,40.00){\makebox(0,0)[cc]{8}}
\put(85.00,70.00){\makebox(0,0)[cc]{6}}
\put(85.00,40.00){\makebox(0,0)[cc]{9}}
\put(55.00,0.00){\makebox(0,0)[cc]{$F_2$}}
\end{picture}
&
\quad
\unitlength 0.50mm
\linethickness{0.4pt}
\begin{picture}(0.00,105.00)
\put(0.00,52.00){\makebox(0,0)[cc]{$\;$}}
\end{picture}
&
\unitlength 0.37mm
\linethickness{0.4pt}
\begin{picture}(110.0,120.00)
\put(10.00,25.00){\framebox(90.33,90.00)[cc]{}}
\put(40.33,115.00){\line(0,-1){90.00}}
\put(70.33,115.00){\line(0,-1){90.00}}
\put(10.00,55.00){\line(1,0){90.33}}
\put(100.33,85.00){\line(-1,0){90.33}}
\put(0.00,40.00){\makebox(0,0)[cc]{$c_2$}}
\put(0.00,70.00){\makebox(0,0)[cc]{$b_2$}}
\put(0.00,100.00){\makebox(0,0)[cc]{$a_2$}}
\put(25.00,125.00){\makebox(0,0)[cc]{$a_1$}}
\put(55.00,125.00){\makebox(0,0)[cc]{$b_1$}}
\put(85.00,125.00){\makebox(0,0)[cc]{$c_1$}}
\put(25.33,15.00){\makebox(0,0)[cc]{$a_1$}}
\put(55.33,15.00){\makebox(0,0)[cc]{$b_1$}}
\put(85.33,15.00){\makebox(0,0)[cc]{$c_1$}}
\put(110.00,40.00){\makebox(0,0)[cc]{$c_2$}}
\put(110.00,70.00){\makebox(0,0)[cc]{$b_2$}}
\put(110.00,100.00){\makebox(0,0)[cc]{$a_2$}}
\put(25.00,100.00){\oval(26.00,26.00)[]}
\put(25.00,100.00){\makebox(0,0)[cc]{1}}
\put(55.00,100.00){\oval(26.00,26.00)[]}
\put(85.00,100.00){\oval(26.00,26.00)[]}
\put(55.00,100.00){\makebox(0,0)[cc]{2}}
\put(85.00,100.00){\makebox(0,0)[cc]{3}}
\put(25.00,70.00){\oval(26.00,26.00)[]}
\put(25.00,40.00){\oval(26.00,26.00)[]}
\put(25.00,70.00){\makebox(0,0)[cc]{4}}
\put(25.00,40.00){\makebox(0,0)[cc]{7}}
\put(55.00,70.00){\oval(26.00,26.00)[]}
\put(55.00,40.00){\oval(26.00,26.00)[]}
\put(85.00,70.00){\oval(26.00,26.00)[]}
\put(85.00,40.00){\oval(26.00,26.00)[]}
\put(55.00,70.00){\makebox(0,0)[cc]{5}}
\put(55.00,40.00){\makebox(0,0)[cc]{8}}
\put(85.00,70.00){\makebox(0,0)[cc]{6}}
\put(85.00,40.00){\makebox(0,0)[cc]{9}}
\put(55.00,0.00){\makebox(0,0)[cc]{$F_1\wedge F_2$}}
\end{picture}
\end{tabular}
 \caption{
Entangled schemes through diagonalization and counterdiagonalization
of the states.
}
\label{2002-statepart2}
\end{figure}
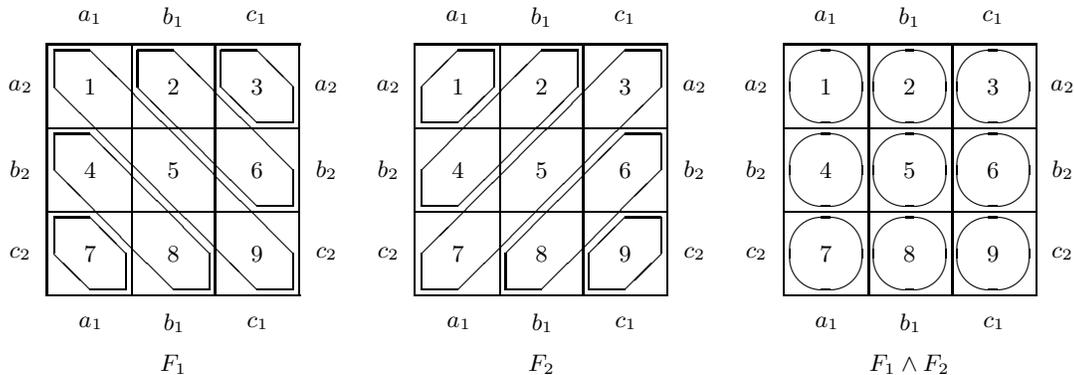

In terms of partitions, entanglement occurs for diagonal
or antidiagonal arrangements of states
which do not add up to completed blocks.
Take, for example, the state partition scheme of Fig. \ref{2002-statepart1},
which results in nonentangled states and state measurements.
A modified, entangled scheme can be established by just grouping the states
into diagonal and counterdiagonal groups as drawn in Fig.
\ref{2002-statepart2}.
The corresponding trits are
\begin{equation}
\begin{array}{llll}
F_1&=&\{\{1,5,9\},\{2,6,7\},\{3,4,8\}\},\\
F_2&=&\{\{1,6,8\},\{2,4,9\},\{3,5,7\}\}.\\
\end{array}
\label{2002-statepart-ps3eentan}
\end{equation}

We can now introduce new $2\times 3$  basis vectors  grouped into the two bases
$\{a_1',b_1',c_1'\}$
and
$\{a_2',b_2',c_2'\}$ by
\begin{equation}
\begin{array}{llll}
\vert a_1'  \rangle &=&{1\over \sqrt{3}}
   (\vert a_1a_2\rangle +  \vert b_1b_2\rangle  + \vert c_1c_2\rangle )
,\\
\vert b_1'  \rangle &=&{1\over \sqrt{3}}
   (\vert a_1b_2\rangle +  \vert b_1c_2\rangle  + \vert c_1a_2\rangle )
,\\
\vert c_1'  \rangle &=&{1\over \sqrt{3}}
   (\vert a_1c_2\rangle +  \vert b_1a_2\rangle  + \vert c_1b_2\rangle )
,\\
\vert a_2'  \rangle &=&{1\over \sqrt{3}}
   (\vert a_1a_2\rangle +  \vert b_1c_2\rangle  + \vert c_1b_2\rangle )
,\\
\vert b_2'  \rangle &=&{1\over \sqrt{3}}
   (\vert a_1b_2\rangle +  \vert b_1a_2\rangle  + \vert c_1c_2\rangle )
,\\
\vert c_2'  \rangle &=&{1\over \sqrt{3}}
   (\vert a_1c_2\rangle +  \vert b_1b_2\rangle  + \vert c_1a_2\rangle )
.
\end{array}
\label{2002-statepart-notb}
\end{equation}
The new orthonormal basis states are ``entangled'' with respect to the old bases
and {\em vice versa}.
Their tensor products generate a complete set of basis states in a new
nine-dimensional Hilbert space.
In terms of the new basis states, the trits can be written as
$F_1\equiv \{\{a_1'\},\{b_1'\},\{c_1'\}\}$
and
$F_2\equiv \{\{a_2'\},\{b_2'\},\{c_2'\}\}$.
The associated bases will be called {\em diagonal bases}.
Note that the permutation which produces the entangled case
(\ref{2002-statepart-ps3eentan})
the nonentangled
(\ref{2002-statepart-ps3e})
one is
$1\rightarrow 1$,
$2\rightarrow 9$,
$3\rightarrow 5$,
$4\rightarrow 6$,
$5\rightarrow 2$,
$6\rightarrow 7$,
$7\rightarrow 8$,
$8\rightarrow 4$,
$9\rightarrow 3$, or $(1)(2,9,3,5)(4,6,7,8)$ in cycle form.
A generalization to diagonal bases associated with
an arbitrary number of nits is straightforward.

In summary we have shown that, by adopting an $n$-ary search strategy,
$k$ particles (entangled or not) specify $k$ nits in such a way that $k$ mutually commuting
measurements of independent observables with $n$ outcomes
are necessary and sufficient to determine the information.
This finding is compatible to Zeilinger's foundational principle for quantum mechanics
\cite{zeil-99}.
In general, the main emphasis in the area of quantum computation
has been in the area of binary decision problems.
It is suggested that these investigations should be extended to
decision problems with $n$ alternatives (e.g., \cite[pp. 332-340]{kleene-52}),
for which quantum information theory seems
to be extraordinarily well equipped.


\end{document}